\documentclass[pre,twocolumn,showpacs,amsmath,amssymb]{revtex4}

\usepackage{amsmath}
\usepackage{amssymb}
\usepackage{graphicx}% Include figure files
\usepackage{dcolumn}% Align table columns on decimal point
\usepackage{bm}% bold math
\usepackage{verbatim}
\usepackage{subfigure}
\usepackage{dcolumn}
\usepackage{xspace}
\usepackage{ifthen}

\newcommand{\pd}{\partial}

\newcommand{\md}{monomer-dimer\xspace}
\newcommand{\dd}{dimer density\xspace}
\newcommand{\p}{\rho}

\newcommand{\two}{true} 

\begin{document}

\title{Exact asymptotics of \md model on rectangular
semi-infinite lattices}
\author{Yong Kong}
\email{matky@nus.edu.sg}
\affiliation{%
Department of Mathematics\\
National University of Singapore\\
Singapore 117543\\
}%

\date{\today}

\begin{abstract}
By using the asymptotic theory of Pemantle and Wilson,
exact asymptotic expansions of the free energy of 
the \md model on rectangular $n \times \infty$ lattices in terms of \dd
are obtained for small values of $n$, 
at both high and low \dd limits.
In the high \dd limit, the theoretical results
confirm the dependence of the free energy on the parity
of $n$, a result obtained previously by computational methods.
In the low \dd limit, 
the free energy on a cylinder $n \times \infty$  lattice strip 
has exactly the same first $n$ terms in the series expansion 
as that of infinite $\infty \times \infty$
lattice.
\end{abstract}


%\keywords{monomer dimer}

\maketitle

\section{Introduction}

In the previous investigation of the \md model on two-dimensional
rectangular lattices by computational methods,
it is found that at high \dd, the free energy depends on the
parity of the width of the 
lattice strip (for a review of \md model, 
see Ref. \onlinecite{Heilmann1970,Kong2006c} and references cited therein).
For a $m \times n$ lattice,
we define the \dd $\p$ as $\p = 2s/mn$, where $s$ is the number of dimers.
With this definition, a close-packed lattice will have $\p=1$.
The grand canonical partition function of the \md system
in a $m \times n$ two-dimensional lattice is
\begin{equation} \label{E:gpf}
Z_{m,n}(x) = \sum_{s=0}^N a_s(m, n) x^s = 
\sum_{0 \le \rho \le 1} a_{m,n}(\rho) x^{mn\rho/2}
\end{equation}
where $a_s(m, n)$ is the number of distinct ways to arrange $s$ dimers
on the $m \times n$ lattice, 
$a_{m,n}(\rho)$ is the corresponding number at the fixed \dd $\p$,
$x$ is the activity of the dimer, and $N$ is the maximum
possible dimer on the lattice: $N = \lfloor mn/2 \rfloor$.
The free energy per lattice site at a given \dd $\rho$ is defined as
\[
 f_{m,n} (\rho) = \frac{\ln a_{m,n}(\rho)}{mn}.
\]

For a semi-infinite $\infty \times n$ lattice at high \dd limit
($\rho \rightarrow 1$),
the computational method shows that the free energy $f_{\infty,n}(\rho)$
depends on the parity of the lattice width $n$ \cite{Kong2006c}:
\begin{equation} \label{E:f_m_n}
 f_{\infty,n}(\rho) \sim
f_{\infty, n}^{\text{lattice}} (1) -  
 \begin{cases}
   (1-\rho) \ln(1-\rho) & \text{$n$ is odd}\\
   \frac{1}{2} (1-\rho) \ln(1-\rho) & \text{$n$ is even}\\
   \end{cases}
\end{equation}
where $f_{\infty, n}^{\text{lattice}} (1)$
is the free energy of close-packed lattice with width $n$, 
the exact expression of which is known and is dependent on the boundary
conditions of the lattice (cylinder or with free edges) 
\cite{Kasteleyn1961,Fisher1961}.
A slightly more general expression, 
for a finite $m \times n$ lattice  at high \dd limit as 
$\rho \rightarrow 1$ and $m \rightarrow \infty$,
is given by \cite{Kong2006c}
\begin{widetext}
\begin{equation} \label{E:f_asympt}
  f_{m, n}(\rho) \sim
  f_{\infty, n}^{\text{lattice}} (1) -
  \left\{ 
  \begin{tabular}{l} 
    $(1-\rho) \ln(1-\rho)$ \\
    $\frac{1}{2} (1-\rho) \ln(1-\rho) $
  \end{tabular} 
  - \frac{1}{2mn} \ln m 
  - \frac{1}{2mn} \ln (1-\rho)
  \quad
  \begin{tabular}{l} 
    $n$ is odd\\
    $n$ is even\\
  \end{tabular} 
  \right.
\end{equation}
\end{widetext}

For the coefficient of the term  $(1-\rho) \ln(1-\rho)$ in the above expansion
of the free energy $f_{\infty, n}(\rho)$,
computational results give values 
very close to the value $-1$ 
when $n$ is odd, and $-1/2$ when $n$ is even.
Although these numerical evidences strongly support the coefficients of
the parity-depending term $(1-\rho) \ln(1-\rho)$ 
as shown in Eq. (\ref{E:f_asympt}), 
there still exists slight possibility that the observations 
be explained in another way.
This is mainly due to the following two reasons.
Firstly, the sequence of $n$ used in the computational studies is not
very long. For cylinder lattices, the largest $n$ used is $17$.
For lattices with free edges, maximum value of $n=16$ is used.
Secondly, at high \dd limit, the convergent rate is the poorest
and some heuristic weighting averages had to be used in the
fitting procedure \cite{Kong2006c}.
Due to these uncertainties, it might be possible that the different
coefficients observed for these small number of 
even and odd lattice widths
are just the initial values of two slowly changing sequences.
When the lattice width $n$ is small,
the deviation of the coefficient from $-1$ or $-1/2$
is so small that it could not be detected numerically. 
As $n$ becomes bigger, these two sequences might converge
to some intermediate values between $-1$ and $-1/2$.

In this report, we use the asymptotic theory of Pemantle and Wilson
\cite{Pemantle2002}
to get exact asymptotics for the free energy $f_{\infty,n}(\rho)$
of $n \times \infty$ lattices.
The exact asymptotics show that the coefficients of  $(1-\rho) \ln(1-\rho)$ 
in the high \dd expansion of $f_{\infty,n}(\rho)$
is exactly $-1$ when $n$ is odd, and $-1/2$ when $n$ is even.
Furthermore, the Pemantle and Wilson method is also used to 
investigate the low \dd expansion of $f_{\infty,n}(\rho)$.
In the low \dd limit, 
the free energy on a cylinder  $n \times \infty$ lattice
has exactly the same first $n$ terms  as that of infinite 
$\infty \times \infty$ lattice. 
In addition, closed form expressions can be obtained
for the differences of the coefficients between 
finite and infinite lattices 
of the next two terms [Eq. (\ref{E:diff})
and Table \ref{T:cylinder_f}].
These properties not only explain the fast convergence rate
of the free energy on cylinder lattices when \dd is low, 
but also provide a quantitative
indicator of the errors when the results of finite size lattices
are used to approximate the infinite lattice.

The article is organized as follows.
In Section \ref{S:PW}, the 
Pemantle and Wilson (PW) method for asymptotics of multivariate
generating functions is summarized.
The starting point for the PW method is the  multivariate
generating function of the model under study, 
and in our case of the \md model,
the generating functions are bivariate.
In Appendix,  
the bivariate generating functions of \md models in two-dimensional 
rectangular lattices are listed for small values of $n$.
These generating functions
are used as the input to the PW method in this article.
In Section \ref{S:high}, the exact asymptotic expansions of
$f_{\infty,n}(\rho)$ at high \dd is derived for some small
values of $n$. The coefficients obtained for the $(1-\rho) \ln(1-\rho)$ term
confirm the dependence on the parity of $n$, as shown 
in Eq. (\ref{E:f_asympt}).
In Section \ref{S:low}, we discuss the exact asymptotic expansions of
$f_{\infty,n}(\rho)$ at low \dd.

\section{Pemantle and Wilson method for asymptotics of multivariate
generating functions} \label{S:PW}

To extract asymptotics from a sequence, it's usually useful to 
utilize its associated generating function.
The method for extract asymptotics from \emph{univariate} generating function
is well-known \cite{Odlyzko95}.
For multivariate generating functions, however, 
the techniques were ``almost entirely missing'' until the
recent development of the Pemantle and Wilson method \cite{Pemantle2002}.
%This theory applies to combinatorial problems when the 
%multivariate generating function of the model is known.
%The method applies to a large class of  
%multivariate generating functions in a systematic way.
For combinatorial problems with known generating functions,
the method can be applied in a systematic way.
The theory applies to generating functions with multiple variables,
and for the bivariate case that we are interested here,
the generating function of two variables takes the form
\begin{equation} \label{E:F}
 G(x, y) = \frac{F(x, y)}{H(x, y)} = \sum_{s, m = 0}^{\infty} a_{sm} x^s y^m
\end{equation}
where $F(x, y)$ and $H(x, y)$ are analytic, and $H(0,0) \ne 0$.
In this case, PW method gives the asymptotic expression
as
\begin{equation} \label{E:asympt}
 a_{sm} \sim \frac{F(x_0, y_0)} {\sqrt{2\pi}} x_0^{-s} y_0^{-m} 
 \sqrt{\frac{-y_0 H_y(x_0, y_0)}{m Q(x_0, y_0)}}
\end{equation}
where $(x_0, y_0)$ is the positive solution to the two equations
%\begin{align} \label{E:asympt_dir}
%  H(x, y)       &= 0 \\
%  s x \frac{ \pd{H} } {\pd {x}}  &= m y \frac{ \pd{H} } {\pd {y}} \notag
%\end{align}
\begin{equation} \label{E:asympt_dir}
  H(x, y)       = 0, \qquad
  m x \frac{ \pd{H} } {\pd {x}}  = s y \frac{ \pd{H} } {\pd {y}} 
\end{equation}
and $Q(x, y)$ is defined as
\ifthenelse{\equal{\two}{true}}{
\begin{align*}
 Q(x, y) =  
 & -(x H_x) (y H_y)^2 -(y H_y) (x H_x)^2 \\
 & - [(y H_y)^2 (x^2 H_{xx}) + (x H_x)^2 (y^2 H_{yy}) \\
 & - 2 (x H_x) (y H_y)(xy H_{xy})]. 
\end{align*}
} {
\[
 -(x H_x) (y H_y)^2 -(y H_y) (x H_x)^2
 - [(y H_y)^2 (x^2 H_{xx}) + (x H_x)^2 (y^2 H_{yy}) 
   - 2 (x H_x) (y H_y)(xy H_{xy})]. 
\]
}
Here $H_x$, $H_y$, etc. are partial derivatives $\pd{H}/\pd{x}$,
$\pd{H}/\pd{y}$, and so on.
One of the advantages of the method over previous ones 
is that the convergence of Eq. (\ref{E:asympt}) is \emph{uniform} 
when $s/m$ and $m/s$ are bounded.

For the \md model discussed here, 
with $n$ as the finite width  of the lattice strip,
$m$ as the length,
and $s$ as the number of dimers, 
we can construct the bivariate generating function $G_n(x, y)$ as
\begin{equation} \label{E:bgf}
 G_n(x, y) = \sum_{m=0}^{\infty} \sum_{s=0}^{mn/2} a_s(m,n) x^s y^m 
 = \sum_{m=0}^{\infty} Z_{m,n}(x) y^m. 
\end{equation}
For the \md model, as well as a large class of lattice models
in statistical physics,
the bivariate generating function $G(x, y)$ 
is always in the form of Eq. (\ref{E:F}),
with $F(x, y)$ and $H(x, y)$ as polynomials in $x$ and $y$.
In fact, for \md model as well as other lattice models,
we can get $H(x, y)$ directly from matrix
$M_n$ used in the recursive formula to calculate the partition functions
\cite{Kong1999,Kong2006,Kong2006b,Kong2006c}
\[
  \Omega_m = M_n \Omega_{m-1}.
\]
Here the vector $\Omega_m$ consists of the partition function Eq. (\ref{E:gpf})
as well as other contracted partition functions \cite{Kong1999}.
The function $H(x, y)$ is closely related to the characteristic 
function of $M_n$ \cite{Kong1999}:
$H(x, y) = \det(M_n - I/y) \times y^w  $, where $w$ is the size of the
matrix $M_n$. More discussions can be found in the Appendix.

The relation between the number of dimers $s$ and the \dd $\p$ is given by
$s = \rho m n /2$.
If we fix the \dd $\p$, 
and substitute this relation into Eq. (\ref{E:asympt_dir}),
then we see that the solution $(x_0, y_0)$ of Eq. (\ref{E:asympt_dir}) 
depends only on $\rho$ and $n$, and does not depend on $m$ or $s$:
\begin{align} \label{E:x0_y0}
  H(x, y)       & = 0, \notag \\
  x \frac{ \pd{H} } {\pd {x}}  &= \frac{\p n y }{2} \frac{ \pd{H} } {\pd {y}} .
\end{align}
Substituting this solution $(x_0(n, \rho), y_0(n, \rho))$ 
into Eq. (\ref{E:asympt})
we obtain
\ifthenelse{\equal{\two}{true}}{
\begin{align} \label{E:asympt_md}
 f_{m,n} (\rho) 
 \sim 
 & -\frac{1}{n} \ln(x_0^{\rho n /2} y_0) 
 -\frac{1}{2}\frac{\ln m}{mn} 
 \notag \\
 & + \frac{1}{mn} \ln \left( F(x_0, y_0) 
\sqrt{\frac{-y_0 H_y(x_0, y_0)}{2\pi Q(x_0, y_0)}}
 \right) .
\end{align}
} {
\begin{equation} \label{E:asympt_md}
 f_{m,n} (\rho) \sim -\frac{1}{n} \ln(x_0^{\rho n /2} y_0) 
 -\frac{1}{2}\frac{\ln m}{mn} 
 + \frac{1}{mn} \ln \left( F(x_0, y_0) 
\sqrt{\frac{-y_0 H_y(x_0, y_0)}{2\pi Q(x_0, y_0)}}
 \right) .
\end{equation}
}
From this asymptotic expansion we obtain the 
logarithmic correction term with coefficient of exactly $-1/2$
(the second term in Eq. (\ref{E:asympt_md})),
for both even and odd values of $n$. 
In fact, the PW asymptotic theory predicts that there exists
such a logarithmic correction term with a coefficient of $-1/2$
for a large class of lattice models when
the two variables involved are proportional, that is, 
when the models are at a fixed ``density''. 
For those lattice models which can be described
by bivariate generating functions,
this logarithmic correction term with a coefficient of $-1/2$
is universal when those models are at  fixed ``density''.
For the \md model, this proportional relation is for
$s$ and $m$ with $s = \rho m n /2$.

When the \dd $\rho$ and the lattice width $n$ are fixed, 
the first term of Eq. (\ref{E:asympt_md}) is
a constant and does not depend on $m$.
We identify it as $f_{\infty,n} (\rho)$
\begin{equation} \label{E:asympt_rho}
 f_{\infty,n} (\rho) = -\frac{1}{n} \ln(x_0^{\rho n /2} y_0) .
\end{equation}

In theory, as long as the bivariate generating function $G(x, y)$
or its denominator $H(x, y)$ is known, 
($x_0$, $y_0$) could be solved from 
the system of equations Eq. (\ref{E:x0_y0}) and 
$f_{\infty,n}(\p)$ could be obtained from Eq. (\ref{E:asympt_rho}).
In practice, however, only very small values of $n$
can be handled this way.
When $n=1$, $x_0$ and $y_0$ can be solved as
\[
 x_0 =  \frac{\p(2-\p)}{4 (1-\p)^2}, \qquad
 y_0 =  \frac{2 (1-\p) }{2-p} 
\]
which leads to
\[
  f_{\infty,1}(\rho) = (1 - \frac{\rho}{2}) \ln (1 - \frac{\rho}{2})
 -\frac{\rho}{2} \ln \frac{\rho}{2}
 - (1-\rho) \ln(1-\rho).
\]
This expression is exact for $0 \le \p \le 1$.

When $n=2$, for both the cylinder lattice and the lattice with
free boundaries, $H(x, y)$ is a cubic polynomial (Appendix).
From Eq. (\ref{E:x0_y0}) a quartic equation can be obtained for $x_0$.
For the cylinder lattice, $x_0$ satisfies the following quartic equation
\ifthenelse{\equal{\two}{true}}{
\begin{multline} \label{E:n2_x0}
32\, \left( 1-\p \right) ^{3}{x}^{4}+144\, \left( 1-\p
 \right) ^{3}{x}^{3}\\
+4\, \left( 1-\p \right)  \left( 10\,{\p}^{2}-20\,\p
+3 \right) {x}^{2} \\
+4\, \left( 2-\p \right)  \left( 3\,{\p}^{2}-3\,\p+1
 \right) x\\
-\p \left( 1-\p \right)  \left( 2-\p \right) = 0 .
\end{multline}
} {
\begin{multline} \label{E:n2_x0}
32\, \left( 1-\p \right) ^{3}{x}^{4}+144\, \left( 1-\p
 \right) ^{3}{x}^{3}+4\, \left( 1-\p \right)  \left( 10\,{\p}^{2}-20\,\p
+3 \right) {x}^{2} \\
+4\, \left( 2-\p \right)  \left( 3\,{\p}^{2}-3\,\p+1
 \right) x-\p \left( 1-\p \right)  \left( 2-\p \right) = 0 .
\end{multline}
}
After $x_0$ is solved, $y_0$ can be obtained as 
a rational function in $\p$ as
\begin{equation} \label{E:n2_y0}
 y_0 = \frac{w(x_0, \p)}{v(\p)}
\end{equation}
where
\begin{widetext}
\begin{multline*}
 w(x_0, \p) = 
\left( 1440\,{\p}^{5}-11424\,{\p}^{2}+12960\,{\p}^{3}-
672+4672\,\p-6976\,{\p}^{4} \right) {x_0}^{3} \\
+ \left( -3360+60432\,{\p}^{3}
-53936\,{\p}^{2}+6608\,{\p}^{5}-32240\,{\p}^{4}+22496\,\p \right) {x_0}^{2} \\
+ \left( 23532\,{\p}^{3}-1716-12072\,{\p}^{4}+2288\,{\p}^{5}-21240\,{\p}^{2
}+9208\,\p \right) x_0 \\
+744\,\p-54+2812\,{\p}^{3}-1552\,{\p}^{4}+300\,{\p}^{5}
-2182\,{\p}^{2}
\end{multline*}
\end{widetext}
and
\[
 v(\p) = \left( 1-\p \right)  \left( 2-\p \right)  \left( 43\,
{\p}^{3}-123\,{\p}^{2}+90\,\p-27 \right) .
\]
Although when $n=2$
closed form expressions could be written down for $x_0$ and $y_0$
(and hence $f_{\infty,2} (\p)$) as a function of $\p$, 
the long expressions are not very informative.
We can, however, obtain highly accurate numerical results from
Eqs. (\ref{E:n2_x0}) and (\ref{E:n2_y0}) for the $n=2$ cylinder lattice   
%(and Eq. (\ref{E:x0_y0}) for lattices with general values of $n$)
 for different values of $\p$.
For example, when $\p = 1/2$, we can solve $x_0$ and $y_0$
numerically from
Eqs. (\ref{E:n2_x0}) and (\ref{E:n2_y0}) as
$x_0 = 
0.389620618156217959$ and
$y_0 = 
0.442004100446556690$, which lead to
$f_{\infty,2} (\frac{1}{2}) 
     = 
0.643863506776659088$.

For the $n=3$ cylinder lattice, in order to solve $x_0$ (or $y_0$),
we have to solve a polynomial equation with a degree of $10$. 
When $\p$ is not so close to $1$, reliable numerical solutions
can be obtained. For example, when $\p = 1/2$,
$x_0$ and $y_0$ can be solved as 
$
0.441361340073863149$ and 
$
0.277272018269763844$ respectively, leading to
$f_{\infty,3} (\frac{1}{2}) = 
0.632058256526951594$.
These ``exact'' numerical values can be used
to check the results obtained previously by the computational methods 
(Table I, Ref. \onlinecite{Kong2006c}).
This procedure confirms the conclusion that the computational methods 
used previously give results with up to $12$ and
$13$ correct digits.

For bigger $n$, it becomes increasingly difficult to solve the system of
polynomial equations Eq. (\ref{E:x0_y0}).
Even numerical solutions become highly unstable, especially at
high \dd.
In the following we investigate the series expansions of
the free energy for lattice strips $n \times \infty$ for small values
of $n$.  
Since the behaviors of the solution 
$x_0$ and $y_0$, and hence the asymptotics of the free energy 
$f_{\infty, n}(\p)$,
are quite different at the high and the low \dd limits,
we discuss the two cases separately.

\section{Asymptotics at the high \dd limit} \label{S:high}

For clarity we define $u = 1 - \p$.
At the high \dd limit when $u \rightarrow 0$,
numerical calculations show that for both odd and even $n$,
$x_0 \rightarrow \infty$ and $y_0 \rightarrow 0$.
If we expand $1/x_0$ and $y_0$ as series of $u$,
from Eq. (\ref{E:x0_y0}) it is found that
$1/x_0$ and $y_0$ have different leading terms
in the series expansion for odd and even $n$.
For odd $n$, the leading term of $1/x_0$ is
$u^2$ and the leading term of $y_0$ is $u^n$.
For even $n$, the leading terms of $1/x_0$ and $y_0$ 
are $u$ and $u^{n/2}$ respectively:
\[
 \frac{1}{x_0} = \sum_{i=2}^\infty a_i u^i, \qquad 
 y_0 = \sum_{i=n}^\infty b_i u^i \qquad 
 \text{when $n$ is odd}
\]
\[
  \frac{1}{x_0} = \sum_{i=1}^\infty a_i u^i, \qquad 
  y_0 = \sum_{i=n/2}^\infty b_i u^i \qquad 
 \text{when $n$ is even}.
\]
These differences in the leading terms of the series expansions
of $x_0$ and $y_0$
lead directly to the
different coefficients of $u \ln u$ in $f_{\infty, n}(u)$
for odd and even $n$. 
By using Eq. (\ref{E:asympt_rho}) we obtain for odd $n$,
\ifthenelse{\equal{\two}{true}}{
\begin{align*}
 f_{\infty, n}(u) 
 & \sim 
  \frac{1-u}{2} \ln (a_2 u^2 + a_3 u^3 + \cdots)  \\
 &  \qquad  - \frac{1}{n} \ln ( b_n u^n + b_{n+1} u^{n+1} + \cdots) \\
 & =    
 \left[ \frac{\ln a_2}{2}  - \frac{\ln b_n}{n}  \right]
 - u \ln u \\
 &  \qquad + \left[ \frac{a_3}{2 a_2} - \frac{b_{n+1}}{n b_n}
   - \frac{\ln a_2}{2}\right] u + \cdots
\end{align*}
} {
\begin{align*}
 f_{\infty, n}(u) 
 & \sim
 \frac{1-u}{2} \ln (a_2 u^2 + a_3 u^3 + \cdots)
 - \frac{1}{n} \ln ( b_n u^n + b_{n+1} u^{n+1} + \cdots) \\
 & = 
 \left[ \frac{\ln a_2}{2}  - \frac{\ln b_n}{n}  \right]
 - u \ln u 
 + \left[ \frac{a_3}{2 a_2} - \frac{b_{n+1}}{n b_n}
   - \frac{\ln a_2}{2}\right] u + \cdots
\end{align*}
}
and for even $n$,
\ifthenelse{\equal{\two}{true}}{
\begin{align*}
 f_{\infty, n}(u) 
 & \sim
 \frac{1-u}{2} \ln (a_1 u + a_2 u^2 + \cdots) \\
 &\qquad  - \frac{1}{n} \ln ( b_{\frac{n}{2}} u^{\frac{n}{2}} 
   + b_{\frac{n}{2}+1} u^{\frac{n}{2}+1} + \cdots) \\
 & = 
 \left[ \frac{\ln a_1}{2}  - \frac{\ln b_{\frac{n}{2}}}{n}  \right]
 - \frac{1}{2} u \ln u \\
 &\qquad  + \left[ \frac{a_2}{2 a_1} - \frac{b_{\frac{n}{2}+1}}{n b_{\frac{n}{2}}}
   - \frac{\ln a_1}{2}\right] u + \cdots
\end{align*}
} {
\begin{align*}
 f_{\infty, n}(u) 
 & \sim
 \frac{1-u}{2} \ln (a_1 u + a_2 u^2 + \cdots) 
 - \frac{1}{n} \ln ( b_{\frac{n}{2}} u^{\frac{n}{2}} 
   + b_{\frac{n}{2}+1} u^{\frac{n}{2}+1} + \cdots) \\
 & = 
 \left[ \frac{\ln a_1}{2}  - \frac{\ln b_{\frac{n}{2}}}{n}  \right]
 - \frac{1}{2} u \ln u 
 + \left[ \frac{a_2}{2 a_1} - \frac{b_{\frac{n}{2}+1}}{n b_{\frac{n}{2}}}
   - \frac{\ln a_1}{2}\right] u + \cdots
\end{align*}
}
The difference in the coefficients of $u \ln u$ in $f_{\infty, n}(u)$
comes directly from the different leading terms of $1/x_0$
for odd and even $n$.

Some explicit expressions of $f_{\infty, n}(\p)$
for cylinder lattices and lattices with free boundaries
are listed in the following.

\subsection{Cylinder lattices}

The asymptotic expansions of $f_{\infty, n}(\p)$ in cylinder lattices
at the high \dd are listed below for $n=1, \dots, 5$. 
For cylinder lattices, the constant term of $f_{\infty, n}(\p)$
is given by the following exact expression \cite{Kasteleyn1961}
\[
  f_{\infty, n}(1) = \frac{1}{n} \ln 
  \prod_{i=1}^{n/2}
  \left[ \sin \frac{(2i - 1)\pi}{n}
  + \left(1 + \sin^2 \frac{(2i - 1)\pi}{n} \right)^{\frac{1}{2}} 
  \right],
\]
which can be used to check 
the constant terms in
the following results.

For $n=1$,
\[
 f_{\infty,1}(u) \sim
 -u \ln u - (\ln2-1) u 
 - \sum_{i=1}^\infty \frac{u^{2i+1}}{2i(2i+1)}.
\]
For $n=2$,
\ifthenelse{\equal{\two}{true}}{
\begin{align*}
f_{\infty, 2}(u) 
   \sim 
   & 
           \frac{1}{2} \,\ln  \left( 1 + \sqrt {2} \right) 
           -\frac{1}{2} u \,\ln \left( u \right) \\
          & +\frac{1}{2} \left[ 1-\ln  \left( 4-2\,\sqrt {2} \right)\right] u \\
          & - \left[ \frac{1}{2} + \frac{1}{8}\,\sqrt {2}  \right] {u}^{2} %\\
           - \left[ \frac{1}{3} - \frac{1}{8}\,\sqrt {2}  \right] {u}^{3} \\
          & - \left[ \frac{1}{3} - {\frac {67}{192}}\,\sqrt {2} \right] {u}^{4} 
	   %\\
           - \left[ \frac{9}{10}- {\frac {45}{64}} \,\sqrt {2} \right] {u}^{5}
	   \\
          & - \left[ \frac {38}{15} - {\frac {6077}{3840}}\,\sqrt {2} \right]
	   {u}^{6} \\
          & - \left[ \frac {142}{21} - {\frac {1169}{256}}\,\sqrt {2} \right]
	   {u}^{7}
%	   - \left[ \frac {134}{7} - {\frac {399103}{28672}}\,\sqrt {2}\right]
%	   {u}^{8}.
\end{align*}
} {
\begin{multline*}
f_{\infty, 2}(u) \sim 
           \frac{1}{2} \,\ln  \left( 1 + \sqrt {2} \right) 
           -\frac{1}{2} u \,\ln \left( u \right)
           +\frac{1}{2} \left[ 1-\ln  \left( 4-2\,\sqrt {2} \right)\right] u
           - \left[ \frac{1}{2} + \frac{1}{8}\,\sqrt {2}  \right] {u}^{2} \\
           - \left[ \frac{1}{3} - \frac{1}{8}\,\sqrt {2}  \right] {u}^{3} %\\
           - \left[ \frac{1}{3} - {\frac {67}{192}}\,\sqrt {2} \right] {u}^{4}
           - \left[ \frac{9}{10}- {\frac {45}{64}} \,\sqrt {2} \right] {u}^{5}
	   \\
           - \left[ \frac {38}{15} - {\frac {6077}{3840}}\,\sqrt {2} \right]
	   {u}^{6} 
           - \left[ \frac {142}{21} - {\frac {1169}{256}}\,\sqrt {2} \right]
	   {u}^{7}
%	   - \left[ \frac {134}{7} - {\frac {399103}{28672}}\,\sqrt {2}\right]
%	   {u}^{8}.
\end{multline*}
}
For $n=3$,
\ifthenelse{\equal{\two}{true}} {
\begin{align*}
  f_{\infty, 3}(u) 
   \sim 
  &
   \frac{1}{6}\,\ln  \left( \frac{5}{2} 
                + \frac{1}{2}\,\sqrt {21} \right)
                - u \,\ln \left( u \right) \\
          &      + \left[ 1-\frac{1}{2}\,\ln  \left( {\frac {6300}{289}}
		  -{\frac {1008}{289}}\,\sqrt {21} \right) \right] u \\
          &      - \left[ {\frac {96}{289}} - {\frac {200}{2023}}\,\sqrt {21} 
		  \right] {u}^{2}\\
	  &	- \left[ {\frac {1975875}{167042}} 
		  - {\frac {1368324}{584647}}\,\sqrt {21} \right] {u}^{3} \\
          &		- \left[ {\frac {2500298208}{24137569}}
		  -{\frac {27592174000}{1182740881}}\,\sqrt {21} \right]{u}^{4}
%		\\
%	  &	- \left[ {\frac {263966005359441}{139515148820}} 
%		  - {\frac {140104720449756}{341812114609}}\,\sqrt {21}
%		  \right] {u}^{5}
%\\
%		- \left[ {\frac {319919000966991072}{10079969502245}}
%		  -{\frac {4798494779367850200}{691485907854007}}\,\sqrt {21} 
%		  \right] {u}^{6} \\
%		- \left[ {\frac {5099295444049416379635}{8156711321216654}}
%		  -{\frac {27249372389950152143364}{199839427369808023}}
%		  \,\sqrt {21} \right] {u}^{7} \\
%		- \left[ 
%		  {\frac {14954643527605183199287584}{1178644785915806503}}
%		  -{\frac {7836316740662406938435253024}
%		    {2829926130983851413703}}\,\sqrt {21} \right] {u}^{8}.
\end{align*}
} {
\begin{multline*}
f_{\infty, 3}(u) \sim 
\frac{1}{6}\,\ln  \left( \frac{5}{2} 
                + \frac{1}{2}\,\sqrt {21} \right)
                - u \,\ln \left( u \right)
                + \left[ 1-\frac{1}{2}\,\ln  \left( {\frac {6300}{289}}
		  -{\frac {1008}{289}}\,\sqrt {21} \right) \right] u
                - \left[ {\frac {96}{289}} - {\frac {200}{2023}}\,\sqrt {21} 
		  \right] {u}^{2}\\
		- \left[ {\frac {1975875}{167042}} 
		  - {\frac {1368324}{584647}}\,\sqrt {21} \right] {u}^{3}
		- \left[ {\frac {2500298208}{24137569}}
		  -{\frac {27592174000}{1182740881}}\,\sqrt {21} \right]{u}^{4}
%		\\
%		- \left[ {\frac {263966005359441}{139515148820}} 
%		  - {\frac {140104720449756}{341812114609}}\,\sqrt {21}
%		  \right] {u}^{5}
%\\
%		- \left[ {\frac {319919000966991072}{10079969502245}}
%		  -{\frac {4798494779367850200}{691485907854007}}\,\sqrt {21} 
%		  \right] {u}^{6} \\
%		- \left[ {\frac {5099295444049416379635}{8156711321216654}}
%		  -{\frac {27249372389950152143364}{199839427369808023}}
%		  \,\sqrt {21} \right] {u}^{7} \\
%		- \left[ 
%		  {\frac {14954643527605183199287584}{1178644785915806503}}
%		  -{\frac {7836316740662406938435253024}
%		    {2829926130983851413703}}\,\sqrt {21} \right] {u}^{8}.
\end{multline*}
}
For $n=4$,
\ifthenelse{\equal{\two}{true}} {
\begin{align*}
f_{\infty, 4}(u) \sim 
       &    -\frac{1}{4}\,\ln  \left( 2-\sqrt {3} \right)
           -\frac{1}{2} u \,\ln \left( u \right) \\
       &   +\frac{1}{2} \left[1 - 
	     \,\ln  \left( {\frac {102}{23}}-{\frac {54}{23}}
	     \,\sqrt {3} \right)  \right] u   \\
       &   - \left[ {\frac {1008}{529}}+{\frac {149}{1058}}\,\sqrt {3} 
	     \right] {u}^{2}\\
       &   - \left[ {\frac {6535949}{839523}}  
	     - {\frac {2581941}{559682}}\,\sqrt {3}
	     \right] {u}^{3} \\
       &	   - \left[ -{\frac {18405319107}{148035889}} 
	     - {\frac {23448085573}{296071778}}\,\sqrt {3} \right] {u}^{4}
%\\
%       &   + \left[ {\frac {316766181828181893989}{2485590672818940}} 
%	     -{\frac {6209832175269970229}{82853022427298}}\,\sqrt {3}
%	     \right] {u}^{5}
%\\
%	   + \left[ {\frac {963519114026534630100844}{986158099440914445}}
%	     -{\frac {178617116081811717717317}{328719366480304815}
%	     }\,\sqrt {3}
%	     \right] {u}^{6} \\
%	   - \left[ {\frac {5575914425782909777147185515}
%	       {1460697376891882475934}}
%	     -{\frac {23785470000185288841071244}
%	       {11592836324538749809}}\,\sqrt {3} \right] {u}^{7} \\
%	   - \left[ {\frac {60502898972656215692651597045947}
%	       {110386987482257975681298}}
%	     -{\frac {122432081130036691496574457474096}
%	       {386354456187902914884543}}\,\sqrt {3} \right] {u}^{8}.
\end{align*}
} {
\begin{multline*}
f_{\infty, 4}(u) \sim 
           -\frac{1}{4}\,\ln  \left( 2-\sqrt {3} \right)
           -\frac{1}{2} u \,\ln \left( u \right)
	   +\frac{1}{2} \left[1 - 
	     \,\ln  \left( {\frac {102}{23}}-{\frac {54}{23}}
	     \,\sqrt {3} \right)  \right] u
	   - \left[ {\frac {1008}{529}}+{\frac {149}{1058}}\,\sqrt {3} 
	     \right] {u}^{2}\\
	   - \left[ {\frac {6535949}{839523}}  
	     - {\frac {2581941}{559682}}\,\sqrt {3}
	     \right] {u}^{3}
	   - \left[ -{\frac {18405319107}{148035889}} 
	     - {\frac {23448085573}{296071778}}\,\sqrt {3} \right] {u}^{4}
%\\
%	   + \left[ {\frac {316766181828181893989}{2485590672818940}} 
%	     -{\frac {6209832175269970229}{82853022427298}}\,\sqrt {3}
%	     \right] {u}^{5}\\
%	   + \left[ {\frac {963519114026534630100844}{986158099440914445}}
%	     -{\frac {178617116081811717717317}{328719366480304815}
%	     }\,\sqrt {3}
%	     \right] {u}^{6} \\
%	   - \left[ {\frac {5575914425782909777147185515}
%	       {1460697376891882475934}}
%	     -{\frac {23785470000185288841071244}
%	       {11592836324538749809}}\,\sqrt {3} \right] {u}^{7} \\
%	   - \left[ {\frac {60502898972656215692651597045947}
%	       {110386987482257975681298}}
%	     -{\frac {122432081130036691496574457474096}
%	       {386354456187902914884543}}\,\sqrt {3} \right] {u}^{8}.
\end{multline*}
}
For $n=5$,
\ifthenelse{\equal{\two}{true}} {
\begin{multline*}
f_{\infty, 5}(u) 
  \sim 
  \frac{1}{5} \ln 
 \frac{\sqrt{5} + \sqrt{41} + \sqrt{15-2\sqrt{5}} + \sqrt{5+2\sqrt{5}} }{4} \\
                 - u \,\ln \left( u \right) \\
  + \left[ 
   1 + \frac{1}{2} \ln 
   \frac{849 + 44 \sqrt{205} + 12 \sqrt{6215 + 422 \sqrt{205}} }{20500}
   \right] u .
\end{multline*}
} {
\begin{multline*}
f_{\infty, 5}(u) \sim 
%\frac{1}{10} \,\ln  \Big[ {\frac {19}{4}}+\frac{1}{4}\,\sqrt {205}
%  +{\frac {39684397}{25377128262444}}\,\sqrt {18683153986535-849133470058
%    \,\sqrt {205}} \\
%  +{\frac {2965859}{25377128262444}}\,\sqrt {205}\sqrt {
%    18683153986535-849133470058\,\sqrt {205}} \Big] 
 \frac{1}{5} \ln 
 \frac{\sqrt{5} + \sqrt{41} + \sqrt{15-2\sqrt{5}} + \sqrt{5+2\sqrt{5}} }{4}
                - u \,\ln \left( u \right) \\
%		+ \left[ 1-\frac{1}{2}\,\ln 
%		  \left( {\frac {1297962604500}{9745440961}}-
%		       {\frac {41794006000}{9745440961}}\,\sqrt {205}
%		       -{\frac {246000}{9745440961}}\,
%		       \sqrt {18683153986535-849133470058\,\sqrt {205}}
%		       \right) 
%		       \right] u
 + \left[ 
   1 + \frac{1}{2} \ln 
   \frac{849 + 44 \sqrt{205} + 12 \sqrt{6215 + 422 \sqrt{205}} }{20500}
   \right] u .
\end{multline*}
}

\subsection{Lattices with free boundaries}
The asymptotic expansions of $f_{\infty, n}(\p)$ in lattices
with free boundaries
at the high \dd are listed below for $n=1, \dots, 4$. 
For lattices with free boundaries, the constant term of $f_{\infty, n}(\p)$
is given by the following exact expression \cite{Kasteleyn1961}
\[
  f_{\infty, n}^{\text{fb}}(1) = \frac{1}{n} \ln \left[ \prod_{i=1}^{\frac{n}{2}}
  \left( \cos \frac{i \pi}{n+1}
  + \left(1 + \cos^2 \frac{i \pi}{n+1} \right)^{\frac{1}{2}} 
  \right) \right],
\]
which can be used to check the constant terms of the following results.

For $n=2$,
\ifthenelse{\equal{\two}{true}} {
\begin{align*}
f_{\infty, 2}^{\text{fb}}(u) \sim 
& \frac{1}{2}\,\ln  \left(\frac{1}{2} + \frac{1}{2}\,\sqrt {5} \right)
-\frac{1}{2} u \,\ln  \left( u \right) \\ 
& + \frac{1}{2} \left[ 1-\ln  \left( 5-2\,\sqrt {5} \right)  \right] u \\
& - \left[ 1+ \frac{1}{20}\,\sqrt {5} \right] {u}^{2}
 + \left[ \frac{2}{5}\,\sqrt {5}-{\frac {13}{12}} \right] {u}^{3} \\
& + \left[ -\frac{8}{3}+{\frac {1073}{600}}\,\sqrt {5} \right] {u}^{4}
 + \left[ {\frac {162}{25}}\,\sqrt {5}-{\frac {561}{40}} \right] {u}^{5} \\
& + \left[ {\frac {221359}{7500}}\,\sqrt {5}-{\frac {1096}{15}} \right] {u}^{6} \\
& - \left[ {\frac {31249}{84}} - {\frac {20564}{125}}\,\sqrt {5} \right] {u}^{7}
%+ \left[ {\frac {13270163}{14000}}\,\sqrt {5}-{\frac {14464}{7}} \right] {u}^{8}
%\\
%+ \left[ -{\frac {1781761}{144}}+{\frac {2082418}{375}}\,\sqrt {5} \right]
%  {u}^{9} .
\end{align*}
} {
\begin{multline*}
f_{\infty, 2}^{\text{fb}}(u) \sim 
\frac{1}{2}\,\ln  \left(\frac{1}{2} + \frac{1}{2}\,\sqrt {5} \right)
-\frac{1}{2} u \,\ln  \left( u \right) 
+ \frac{1}{2} \left[ 1-\ln  \left( 5-2\,\sqrt {5} \right)  \right] u
- \left[ 1+ \frac{1}{20}\,\sqrt {5} \right] {u}^{2}\\
+ \left[ \frac{2}{5}\,\sqrt {5}-{\frac {13}{12}} \right] {u}^{3}
+ \left[ -\frac{8}{3}+{\frac {1073}{600}}\,\sqrt {5} \right] {u}^{4}
+ \left[ {\frac {162}{25}}\,\sqrt {5}-{\frac {561}{40}} \right] {u}^{5} \\
+ \left[ {\frac {221359}{7500}}\,\sqrt {5}-{\frac {1096}{15}} \right] {u}^{6}
- \left[ {\frac {31249}{84}} - {\frac {20564}{125}}\,\sqrt {5} \right] {u}^{7}
+ \left[ {\frac {13270163}{14000}}\,\sqrt {5}-{\frac {14464}{7}} \right] {u}^{8}
\\
+ \left[ -{\frac {1781761}{144}}+{\frac {2082418}{375}}\,\sqrt {5} \right]
  {u}^{9} .
\end{multline*}
}
For $n=3$,
\ifthenelse{\equal{\two}{true}} {
\begin{align*}
f_{\infty, 3}^{\text{fb}}(u) \sim 
& \frac{1}{6}\,\ln  \left( 2+\sqrt {3} \right)
- u \ln  \left( u \right) \\
& + \left[ 1 +\ln  \left( \frac{1}{36}\,\sqrt {6}+ \frac{1}{6}\,\sqrt {2} 
  \right)  \right] u \\
& + \left[ {\frac {103}{121}}\,\sqrt {3}+{\frac {240}{121}} \right] {u}^{2}\\
& - \left[ {\frac {112740}{
14641}}\,\sqrt {3}+{\frac {806673}{29282}} \right] {u}^{3} \\
& + \left[ {\frac {369777941}{1771561}}\,\sqrt {3}+{
\frac {492403464}{1771561}} \right] {u}^{4}\\
& - \left[ {\frac {
30323479269681}{4287177620}}+{\frac {662980595688}{214358881}}\,\sqrt
{3} \right] {u}^{5}
%\\
%+ \left[
%{\frac {11177133877193769}{129687123005}}\,\sqrt {3}+{\frac {
%3274443199857600}{25937424601}} \right] {u}^{6}\\
%- \left[ {\frac {14126895904648156227}{3994363388554}}+{\frac {
%514260738911599164}{285311670611}}\,\sqrt {3} \right] {u}^{7}\\
%+ \left[ {\frac {
%19808456894016122941848}{241658985007517}}+{\frac {
%12433205045432130373233}{241658985007517}}\,\sqrt {3} \right] {u}^{8} .
%-\left[ {\frac {79125321214382847472498059}{
%33417985355325208}}+{\frac {5386054341184656542431668}{
%4177248169415651}}\,\sqrt {3} \right] {u}^{9}
\end{align*}
} {
\begin{multline*}
f_{\infty, 3}^{\text{fb}}(u) \sim 
\frac{1}{6}\,\ln  \left( 2+\sqrt {3} \right)
- u \ln  \left( u \right)
+ \left[ 1 +\ln  \left( \frac{1}{36}\,\sqrt {6}+ \frac{1}{6}\,\sqrt {2} 
  \right)  \right] u
+ \left[ {\frac {103}{121}}\,\sqrt {3}+{\frac {240}{121}} \right] {u}^{2}\\
- \left[ {\frac {112740}{
14641}}\,\sqrt {3}+{\frac {806673}{29282}} \right] {u}^{3}
+ \left[ {\frac {369777941}{1771561}}\,\sqrt {3}+{
\frac {492403464}{1771561}} \right] {u}^{4}\\
- \left[ {\frac {
30323479269681}{4287177620}}+{\frac {662980595688}{214358881}}\,\sqrt
{3} \right] {u}^{5}
%\\
%+ \left[
%{\frac {11177133877193769}{129687123005}}\,\sqrt {3}+{\frac {
%3274443199857600}{25937424601}} \right] {u}^{6}\\
%- \left[ {\frac {14126895904648156227}{3994363388554}}+{\frac {
%514260738911599164}{285311670611}}\,\sqrt {3} \right] {u}^{7}\\
%+ \left[ {\frac {
%19808456894016122941848}{241658985007517}}+{\frac {
%12433205045432130373233}{241658985007517}}\,\sqrt {3} \right] {u}^{8} .
%-\left[ {\frac {79125321214382847472498059}{
%33417985355325208}}+{\frac {5386054341184656542431668}{
%4177248169415651}}\,\sqrt {3} \right] {u}^{9}
\end{multline*}
}
For $n=4$,
\ifthenelse{\equal{\two}{true}} {
\begin{align*}
f_{\infty, 4}^{\text{fb}}(u) 
& \sim 
 \frac{1}{4} \ln \frac{\sqrt{5} + 1 + \sqrt{22 + 2\sqrt{5}} }{4} \\
&
+\frac{1}{4} \ln \frac{\sqrt{5} - 1 + \sqrt{22 - 2\sqrt{5}} }{4} \\
&
-\frac{1}{2} u \,\ln  \left( u \right)  \\
&
+\biggl[ 
  \frac{1}{2} + \frac{1}{2}\,\ln  \frac{341801}{2} \\
& \qquad
 -\frac{1}{2}\,\ln  \biggl( 545403+81734\,\sqrt
{29} \\
& \qquad \qquad -4\,\sqrt {27680943526+5123717738\,\sqrt {29}} \biggr)
\biggr] u .
\end{align*}
} {
\begin{multline*}
f_{\infty, 4}^{\text{fb}}(u) \sim 
%\frac{1}{4} \,\ln  \left[
%  \frac{1+\sqrt {29}}{4}
%  +{\frac {4873 \sqrt {
%	954515294+176679922\,\sqrt {29}}}{34180100}} 
%  -{\frac {181 \sqrt {29}
%      \sqrt {954515294+176679922\,\sqrt {29}}}{8545025}} 
%  \right] \\
 \frac{1}{4} \ln \frac{\sqrt{5} + 1 + \sqrt{22 + 2\sqrt{5}} }{4}
+\frac{1}{4} \ln \frac{\sqrt{5} - 1 + \sqrt{22 - 2\sqrt{5}} }{4}
-\frac{1}{2} u \,\ln  \left( u \right)  \\
+\left[ 
  \frac{1}{2} + \frac{1}{2}\,\ln  \frac{341801}{2} 
 -\frac{1}{2}\,\ln  \left( 545403+81734\,\sqrt
{29}-4\,\sqrt {27680943526+5123717738\,\sqrt {29}} \right)
\right] u .
\end{multline*}
}

\section{Asymptotics at the low \dd limit} \label{S:low}

Unlike the high \dd case, at low \dd when $\p \rightarrow 0$,
numerical calculations  show  that $x_0$ approaches zero
and $y_0$ approaches 1, for both odd and even values of $n$.
The series expansions of $x_0$ and $y_0$ thus have the following forms:
\[
 x_0 = \sum_{i=1}^\infty a_i \p^i, \qquad 
 y_0 = 1 + \sum_{i=1}^\infty b_i \p^i
\]
From Eq. (\ref{E:asympt_rho}), 
the general form of the free energy at the low \dd is 
\ifthenelse{\equal{\two}{true}} {
\begin{align*}
 f_{\infty, n}(\p) 
  \sim 
 & -\frac{\p \ln \p }{2} 
 -\left[
   \frac{\ln a_1}{2} + \frac{b_1}{2}
   \right] \p \\
 & -\left[
   \frac{a_2}{2 a_1} + \frac{b_2}{n} - \frac{b_1^2}{2n}
   \right] \p^2 + \cdots
\end{align*}
} {
\[
 f_{\infty, n}(\p) 
  \sim 
 -\frac{\p \ln \p }{2} 
 -\left[
   \frac{\ln a_1}{2} + \frac{b_1}{2}
   \right] \p
 -\left[
   \frac{a_2}{2 a_1} + \frac{b_2}{n} - \frac{b_1^2}{2n}
   \right] \p^2 + \cdots
\]
}
For both odd and even values of $n$, at low \dd the coefficient
of logarithmic term $\p \ln \p$ is $-1/2$, consistent with the previous 
results obtained by computational methods
\cite{Kong2006b}.
This coefficient comes directly from the fact that the leading term
in the series expansion of $x_0$ is $a_1 \p$.

The asymptotic expansions of free energy in cylinder lattices
at low \dd show an interesting property: for lattice strip $n \times \infty$
with a width of $n$, the first $n$ terms in the series expansion 
of $f_{\infty, n}(\p)$
is exactly the same as the first $n$ terms in the series expansion 
of $f_{\infty, \infty}(\p)$,
the free energy of the infinite lattice. 
In order to compare the free energy in semi-infinite $n \times \infty$
lattice strips 
with that of an infinite $\infty \times \infty$ lattice,
in the following section the series of Gaunt \cite{Gaunt1969} is used
to derive the series of  $f_{\infty, \infty}(\p)$.

\subsection{Free energy for an infinite lattice}

Gaunt gave a series expansion of the dimer activity $x$ as a function of
the number density $t$  (Ref. \onlinecite{Gaunt1969},
column 2 of Table II):
\ifthenelse{\equal{\two}{true}} {
\begin{align} \label{E:gaunt_s}
x(t) = 
& t+7\,{t}^{2}+40\,{t}^{3}+206\,{t}^{4} 
+1000\,{t}^{5} \notag \\
&+4678\,{t}^{6} %\\
 +21336\,{t}^{7} %\\
+95514\,{t}^{8} \notag \\
& +421472\,{t}^{9}+1838680\,{t}^{10} 
 +7947692\,{t}^{11} \notag \\
&+34097202\,{t}^{12}+145387044\,{t}^{13} \notag \\
&+616771148\,{t}^{14}+2605407492\,{t}^{15} + \cdots.
\end{align}
} {
\begin{multline} \label{E:gaunt_s}
x(t) = 
t+7\,{t}^{2}+40\,{t}^{3}+206\,{t}^{4} 
+1000\,{t}^{5}+4678\,{t}^{6}+21336\,{t}^{7} %\\
+95514\,{t}^{8} \\
+421472\,{t}^{9}+1838680\,{t}^{10} 
+7947692\,{t}^{11}+34097202\,{t}^{12}+145387044\,{t}^{13} \\
+616771148\,{t}^{14}+2605407492\,{t}^{15} + \cdots.
\end{multline}
}
Here $t = \theta/4$, where $\theta$
is the average number of sites covered by dimers
when grand canonical ensembles are considered \cite{Kong2006c}.

By using the relation between the grand canonical ensemble
and the canonical ensemble \cite{Kong2006c} and identifying $\theta$
with the \dd $\p$, we have the following relation
\[
f_{\infty, \infty}(\p) = 
-\frac{1}{2} \int  \ln \left[ x \left( \frac{\p}{4} \right) \right]  d\p.
\]
From this relation and the series in Eq. (\ref{E:gaunt_s}) we can obtain 
the series expression for the free energy of the infinite lattice
\ifthenelse{\equal{\two}{true}} {
\begin{align} \label{E:gaunt}
f_{\infty, \infty}(\p)=
& -\frac{1}{2}\,\p\ln  \left( \p \right)
+\left[ \frac{1}{2} + \ln  \left( 2 \right) \right] \p
-{\frac {7}{16}}\,{\p}^{2}  \notag \\
& -{\frac {31}{192}}\,{\p}^{3}
-{\frac {121}{1536}}\,{\p}^{4}-{\frac {471}{10240}}\,{\p}^{5} \notag \\
& 
-{\frac {1867}{61440}}\,{\p}^{6} 
-{\frac {7435}{344064}}\,{\p}^{7}
-{\frac {4211}{262144}}\,{\p}^{8}  \notag \\
&
-{\frac {116383}{9437184}}\,{\p}^{9}
-{\frac {459517}{47185920}}\,{\p}^{10}
-{\frac {1821051}{230686720}}\,{\p}^{11} \notag \\
&
-{\frac {7255915}{1107296256}}\,{\p}^{12}
-{\frac {9687973}{1744830464}}\,{\p}^{13} \notag \\
&
-{\frac {16697149}{3489660928}}\,{\p}^{14} \notag \\
&
-{\frac {157001097}{37580963840}}\,{\p}^{15} + \cdots .
\end{align}
} {
\begin{multline} \label{E:gaunt}
f_{\infty, \infty}(\p)=
-\frac{1}{2}\,\p\ln  \left( \p \right)
+\left[ \frac{1}{2} + \ln  \left( 2 \right) \right] \p
-{\frac {7}{16}}\,{\p}^{2}-{\frac {31}{192}}\,{\p}^{3}
-{\frac {121}{1536}}\,{\p}^{4}-{\frac {471}{10240}}\,{\p}^{5}
-{\frac {1867}{61440}}\,{\p}^{6} \\
-{\frac {7435}{344064}}\,{\p}^{7}
-{\frac {4211}{262144}}\,{\p}^{8}
-{\frac {116383}{9437184}}\,{\p}^{9}
-{\frac {459517}{47185920}}\,{\p}^{10}
-{\frac {1821051}{230686720}}\,{\p}^{11} \\
-{\frac {7255915}{1107296256}}\,{\p}^{12}
-{\frac {9687973}{1744830464}}\,{\p}^{13}
-{\frac {16697149}{3489660928}}\,{\p}^{14}
-{\frac {157001097}{37580963840}}\,{\p}^{15} + \cdots .
\end{multline}
}

\subsection{Cylinder lattices}

The coefficients of $\p^i$ in the series expansion
of $f_{\infty, n}(\p)$ at low \dd is listed in Table \ref{T:cylinder_f}
for $n=1, \dots, 7$.
The term $-\frac{1}{2} \p \ln(\p)$, which is common to lattices of all sizes,
is not included in the Table.
Also listed in the last row of the Table 
are the coefficients for the infinite lattice (Eq. (\ref{E:gaunt})).
It is evident from the Table that 
for lattice strip $n \times \infty$
with a width of $n$, the first $n$ terms of $f_{\infty, n}(\p)$
(including the term of $-\frac{1}{2} \p \ln(\p)$)
is exactly the same as the first $n$ terms 
of the infinite lattice $f_{\infty, \infty}(\p)$.
For example, the lattice strip $7 \times \infty$
has the first seven terms identical to those of the infinite lattice,
up to the term of $\p^6$.
This nice property gives a quantitative estimate of the error
when we use the values of finite lattices to approximate
the properties of the infinite lattice.  
It also explains why the sequence of the free energy in cylinder lattices
converges so fast, especially when $\p$ is small \cite{Kong2006c}. 

The term of $\p^n$ in $f_{\infty, n}(\p)$
is the first term that differs from the series expansion of 
$f_{\infty, \infty}(\p)$.
The difference between the coefficients of $\p^n$ in $f_{\infty, n}(\p)$
and $f_{\infty, \infty}(\p)$
shows a regular pattern:
starting from $n=2$, the differences are
$\frac{1}{32}, -\frac{1}{192}, \frac{1}{1024}, 
-\frac{1}{5120}, \frac{1}{24576}, -\frac{1}{114688}, \dots$.
For example, for $n=2$, $-\frac{13}{32} - (-\frac{7}{16}) = \frac{1}{32}$.
The closed form of this alternating sequences
clearly is $\frac{(-1)^{n}}{n 4^n}$.

The difference between the coefficients of $\p^{n+1}$,
the second term that differs between
finite and  infinite lattices, also shows a regular pattern.
Starting from $n=3$, the sequence is
$-\frac{1}{256}, \frac{1}{1024}, 
-\frac{1}{4096}, \frac{1}{16384},
-\frac{1}{65536}, \dots $. It is obvious that the sequence
takes the closed form expression $\frac{(-1)^{n}}{4 \cdot 4^n}$.
Due to the limited number of data points in Table \ref{T:cylinder_f},
currently it is not clear whether the differences of higher degree terms 
also can be written in simple closed forms.

From the closed form expressions of these two dominant terms
that differ between finite and infinite lattices
at low \dd,
we see that when $n \ge 3$
\begin{equation} \label{E:diff}
 f_{\infty, \infty}(\p)   \sim 
 f_{\infty, n}(\p)
 -\frac{(-1)^n}{4^n} 
 \left( \frac{\p^n}{n} 
 +  \frac{\p^{n+1}}{4} \right)
 + O(\p^{n+2}).
\end{equation}

The coefficients of $\p^i$ in the series expansion
of $x_0$ and $-(\ln y_0) / n$ are also listed in Tables 
\ref{T:cylinder_x0} and \ref{T:cylinder_lny0n}.
From the Tables we can see that for cylinder lattice strips
$n \times \infty$, $x_0$ and $(\ln y_0) / n$ share the same first
$n-1$ terms with their corresponding series expansions
of the infinite lattice.

\begingroup
\squeezetable
\begin{table*}
\caption{Coefficients of $\p^i$ in the series expansion
of $f_{\infty, n}(\p)$ of cylinder lattices at low \dd.  
The term $-\frac{1}{2} \p \ln(\p)$,
common to lattices of all sizes, is not shown. 
The last row for the infinite
lattice $f_{\infty, \infty}(\p)$ 
is taken from the series expansion Eq. (\ref{E:gaunt}).
Terms of $f_{\infty, n}(\p)$
that are equal to those of the infinite lattice are underlined.
\label{T:cylinder_f}}

\begin{ruledtabular}
%\begin{tiny}
\begin{tabular}{ccccccccc}

$n$ & 
$\p$ &
  $-\p^2$ &  $-\p^3$ &  $-\p^4$ &  $-\p^5$ &  $-\p^6$ &  $-\p^7$ &  $-\p^8$
 \\
% &  $\p^9$ \\

\hline

1 &
$\frac{\ln2+1}{2}$ &
3/8  &  7/48  &  5/64  &  31/640  &  21/640  &  127/5376  &  255/14336  \\
%&  511/36864 \\
%\cline{2-2}

2 &
$\ln2 + 1/2$ &
13/32  &  115/768  &  419/6144  &  5491/163840  &  17489/983040  &  116687/11010048  &  423771/58720256  \\
%&  13223491/2415919104 \\
\cline{2-2}

3 &
$\ln2 + 1/2$ &
7/16  &  1/6  &  127/1536  &  1027/20480  &  8653/245760  &  17677/688128  &  2381/131072 \\
% &  932485/75497472 \\
\cline{2-3}

4 &
$\ln2 + 1/2$ &
7/16  &  31/192  &  239/3072  &  461/10240  &  3569/122880  &  27325/1376256  &  7579/524288  \\
%&  52877/4718592 \\
\cline{2-4}

5 &
$\ln2 + 1/2$ &
7/16  &  31/192  &  121/1536  &  473/10240  &  941/30720  &  3791/172032  &  4375/262144  \\
%&  7657/589824\\
\cline{2-5}

6 &
$\ln2 + 1/2$ &
7/16  &  31/192  &  121/1536  &  471/10240 & 
1243/40960 &
3707/172032 &
4177/262144  \\
%& 114391/9437184\\
\cline{2-6}

7 &
$\ln2 + 1/2$ &
7/16 &
31/192 &
121/1536 &
471/10240 &
1867/61440 &
3719/172032 &
4215/262144 \\
\cline{2-7}

$\infty$ &
$\ln2 + 1/2$ &
7/16  &  31/192  &  121/1536  & 471/10240 &
1867/61440  &
7435/344064 &
4211/262144 
%&116383/9437184
\end{tabular}
%\end{tiny}
\end{ruledtabular}
\end{table*}
\endgroup

\begingroup
\squeezetable
\begin{table*}
\caption{The coefficients in the series expansion of
$x_0$ for cylinder lattice strips $n \times \infty$ at low \dd. 
Terms that are equal to those of the infinite lattice are underlined.
\label{T:cylinder_x0}}

\begin{ruledtabular}
%\begin{tiny}
\begin{tabular}{ccccccccc}

$n$ & 
$\p$ &  $\p^2$ &  $\p^3$ &  $\p^4$ &  $\p^5$ &  $\p^6$ &  $\p^7$ &  $\p^8$
 \\
% & $\p^9$ \\

\hline

1 &
1/2  &  3/4    &  1       &  5/4        &  3/2        &  
   7/4         &  2       &  9/4      \\
%  &  5/2        \\
  
2 &
1/4  &  13/32  &  71/128  &  1393/2048  &  6353/8192  &  
   55073/65536  &  230343/262144  &  7519577/8388608  \\
%&  30181645/33554432  \\
\cline{2-2}

3 &
1/4  &  7/16   &  81/128  &  423/512    &  4179/4096  &  
   9993/8192  &  23341/16384  &  854147/524288  \\
%&  7661421/4194304  \\
\cline{2-3}

4 &
1/4  &  7/16   &  5/8  &  411/512       &  497/512    &  
   289/256  &  20917/16384  &  370861/262144  \\
%&  404955/262144  \\
\cline{2-4}

5 &
1/4  &  7/16   &  5/8  &  103/128       &  2001/2048  &  
   9369/8192  &  42803/32768  &  192145/131072  \\
%&  1701937/1048576  \\
\cline{2-5}

6 &
1/4  &  7/16   &  5/8  &  103/128       &  125/128    &  9355/8192  &
   21329/16384 & 190871/131072 \\
%& 210371/131072 \\
\cline{2-6}

7 &

1/4  &  7/16   &  5/8  &  103/128       &  125/128   & 2339/2048 &
   42673/32768 & 191043/131072          \\
%&  421565/262144\\
\cline{2-7}

$\infty$ &
1/4  &  7/16   &  5/8  &  103/128       &  125/128    &
  2339/2048 &
%  2667/2048(?) 

\end{tabular}
%\end{tiny}
\end{ruledtabular}
\end{table*}
\endgroup

\begingroup
\squeezetable
\begin{table*}
\caption{The coefficients in the series expansion of
$-\ln(y_0)/n$ for cylinder lattice strips $n \times \infty$ at low \dd. 
Terms that are equal to those of the infinite lattice are underlined.
\label{T:cylinder_lny0n}}

\begin{ruledtabular}
%\begin{tiny}
\begin{tabular}{ccccccccc}

$n$ & 
$\p$ &  $\p^2$ &  $\p^3$ &  $\p^4$ &  $\p^5$ &  $\p^6$ &  $\p^7$ &  $\p^8$
\\
%  &  $\p^9$ \\

\hline

1 &
1/2  &  3/8  &  7/24  &  15/64  &  31/160  &  21/128  &  127/896  &  255/2048
\\ 
% &  511/4608  \\
\cline{2-2}
2 &
1/2  &  13/32  &  115/384  &  419/2048  &  5491/40960  &  17489/196608  &  116687/1835008  &  423771/8388608 \\
% &  13223491/301989888  \\
\cline{2-2}

3 &
1/2  &  7/16  &  1/3  &  127/512  &  1027/5120  &  8653/49152  &  17677/114688  &  16667/131072  \\
%&  932485/9437184 \\
\cline{2-3}

4 &
1/2  &  7/16  &  31/96  &  239/1024  &  461/2560  &  3569/24576  &  27325/229376  &  53053/524288  \\
%&  52877/589824 \\
\cline{2-4}

5 &
1/2  &  7/16  &  31/96  &  121/512  &  473/2560  &  941/6144   &  3791/28672  &  30625/262144  \\
%&  7657/73728 \\
\cline{2-5}

6 &
1/2 &   7/16  &  31/96  &  121/512  &  471/2560  &  1243/8192  &  3707/28672  &
29239/262144  \\
%&  114391/1179648\\
\cline{2-6}

7 &
1/2 &   7/16  &  31/96  &  121/512  &  471/2560  &  1867/12288 & 3719/28672   &
29505/262144  \\
\cline{2-7}

$\infty$ &
1/2  &  7/16  &  31/96  &  121/512  &  471/2560  &  1867/12288 & 
%7435/57344(?) 

\end{tabular}
%\end{tiny}
\end{ruledtabular}
\end{table*}
\endgroup

\subsection{Lattices with free boundaries}

The coefficients of $\p^i$ in the series expansion
of $f_{\infty, n}(\p)$ for lattices with free boundaries at low \dd 
is listed in Table \ref{T:fb_f}
for $n=2, 3$, and $4$.
The term $-\frac{1}{2} \p \ln(\p)$, which is common to lattices of all sizes,
is not included in the Table.
From the Table we can see that,
unlike cylinder lattices, 
none of the coefficients 
is the same as the coefficients of the infinite lattice.

As for the cylinder lattices,
the coefficients of $\p^i$ in the series expansion
of $x_0$ and $-(\ln y_0) / n$ are also listed in Tables 
\ref{T:fb_x0} and \ref{T:fb_lny0n}.
Again, except for the coefficient of $\p$ in $(\ln y_0) / n$,
none of the coefficients 
is the same as the coefficients of the infinite lattice.

\begingroup
\squeezetable
\begin{table*}
\caption{Coefficients of $\p^i$ in the series expansion
of $f_{\infty, n}(\p)$ of lattices with free boundaries at low \dd.  
The term $-\frac{1}{2} \p \ln(\p)$,
common to lattices of all sizes, is not shown. 
Numbers in square brackets denote powers of $10$.
\label{T:fb_f}}

\begin{ruledtabular}
\begin{tiny}
\begin{tabular}{ccccccccc}
$n$ & 
$\p$ &
  $-\p^2$ &  $-\p^3$ &  $-\p^4$ &  $-\p^5$ &  $-\p^6$ &  $-\p^7$ &  $-\p^8$
 \\
% &  $\p^9$ \\

\hline

2 &
$\frac{1}{2} (\ln 3+1) $  & %1
5/12                 & %2           0.4167     d<
17/108               & %3           0.1574     u<
49/648               & %4           0.0756     u<
403/9720             & %5           0.04146    d<
25/972               & %6           0.02572    d<
3223/183708          & %7           0.017544   d<
27569/2204496        \\%8           0.0125     d<

3 &
$\frac{1}{2} (\ln (10/3)+1) $  & %1
87/2[2]                     & %2    0.435      d< 0.4375 = 7/16
1569/1[4]                 & %3      0.1569     u< 0.16145833 = 31/192
14697/2[5]               & %4       0.073485   d< 0.078776 = 121/1536
2208627/5[7]           & %5         0.04417254 d< 0.045996 = 471/10240
192832569/625[7]       & %6         0.0308532  u> 0.030387 = 1867/61440
9382406661/4375[8]    & %7          0.0214455  u< 0.0216 = 7435/344064
1278517726503/875[11] \\ %8         0.01461    d< 0.016 = 4211/262144

4 &
$\frac{1}{2}(\ln (7/2)+1) $  & %1
43/98                     & %2      0.43877551  > 0.4375 = 7/16
1123/7203                 & %3      0.1559      < 0.16145833 = 31/192
26323/352947              & %4      0.07458     < 0.078776 = 121/1536
187546/4117715            & %5      0.045546    < 0.045996 = 471/10240
18361592/605304105        & %6      0.030334    < 
850524016/41523861603     & %7      0.02048     <
70713460952/4747561509943        %  0.01489469  <

\end{tabular}
\end{tiny}
\end{ruledtabular}
\end{table*}
\endgroup

\begingroup
\squeezetable
\begin{table*}
\caption{$x_0$
The coefficients in the series expansion of
$x_0$ for $n \times \infty$ lattices with free boundaries at low \dd. 
Numbers in square brackets denote powers of $10$.
\label{T:fb_x0}}

\begin{ruledtabular}
\begin{tiny}
\begin{tabular}{ccccccccc}

$n$ & 
$\p$ &  $\p^2$ &  $\p^3$ &  $\p^4$ &  $\p^5$ &  $\p^6$ &  $\p^7$ &  $\p^8$
 \\
% & $\p^9$ \\
\hline

2 &
1/3 &
5/9 &
7/9 &
239/243 &
851/729 &
2909/2187 &
29017/19683 &
31507/19683 \\

3 &
3/10     &
261/5[2]  &
9207/125[2] &
116397/125[3] &
4826031/3125[4] &
021574591/15625[5] &
28645564383/1953125[4] &
3178995484149/1953125[6] \\

4 &
2/7     &
172/343 &
11888/16807 &
738720/823543 &
6219872/5764801 &
354308800/282475249 &
137557926784/96889010407 &
7481262371584/4747561509943

\end{tabular}
\end{tiny}
\end{ruledtabular}
\end{table*}
\endgroup

\begingroup
\squeezetable
\begin{table*}
\caption{$-\ln(y_0)/n$
The coefficients in the series expansion of
$-(\ln y_0)/n$ for $n \times \infty$ lattices with free boundaries at low \dd. 
Numbers in square brackets denote powers of $10$.
\label{T:fb_lny0n}}

\begin{ruledtabular}
\begin{tiny}
\begin{tabular}{ccccccccc}

$n$ & 
$\p$ &  $\p^2$ &  $\p^3$ &  $\p^4$ &  $\p^5$ &  $\p^6$ &  $\p^7$ &  $\p^8$
\\
%  &  $\p^9$ \\

\hline

2 &
1/2   &
5/12  &
17/54  &
49/216 &
403/2430 &
125/972 &
3223/30618 &
27569/314928  \\

3 &
1/2  &
87/2[2]  &
1569/5[3]  &
44091/2[5]  &
2208627/125[5]  &
192832569/125[7]  &
28147219983/21875[7] &
1278517726503/125[11] \\

4 &
1/2  &
43/98  &
2246/7203  &
26323/117649 &
750184/4117715 &
18361592/121060821 &
1701048032/13841287201 &
70713460952/678223072849

\end{tabular}
\end{tiny}
\end{ruledtabular}
\end{table*}
\endgroup

\appendix

\section{Bivariate generating functions of \md models in two-dimensional 
planar lattices} \label{S:GF}

The bivariate generating functions of \md models
can be derived directly from the matrices $M_n$ which are used
in the computational studies \cite{Kong1999,Kong2006,Kong2006b,Kong2006c}.
The denominator $H(x, y)$ of the bivariate generating functions
$G(x, y)$ is directly related to the characteristic function of $M_n$: 
\[
 H(x, y) = \det(M_n - y^{-1} I) \times y^w,
\]
where $w$ is the size of the matrix $M_n$.
For the purpose of this article, it is sufficient to know the information
of $H(x, y)$.
From $H(x, y)$ and a few initial terms, the bivariate generating functions
can be obtained. 
For completeness we list $G(x, y)$ for small values of $n$ in this Appendix.
The cylinder lattices and lattices with free edges
are listed separately.
Variable $x$ is associated with the number of dimers $s$ and
variable $y$ is associated with the length of the lattice $m$,
as defined in Eq. (\ref{E:bgf}).

When $n=1$, there is no distinction between these two boundary conditions:
\[
 G_1 = \frac{1}{1-y-xy^2} .
\]

\subsection{Cylinder lattices}
For $n=2$,
\[
{G_2^c}={\frac {1-yx}{1- \left( 1+3\,x \right) y+x \left( x-1
 \right) {y}^{2}+{y}^{3}{x}^{3}}} .
\]
For $n=3$,
\begin{widetext}
\[
{G_3^c}={\frac {1-2\,yx-{y}^{2}{x}^{3}}{{x}^{6}{y}^{4}-{x}^{3}
 \left( x-1 \right) {y}^{3}-x \left( 1+3\,x+5\,{x}^{2} \right) {y}^{2}
- \left( 5\,x+1 \right) y+1}} .
\]
\end{widetext}
For $n=4$,
$
 {G_4^c}=F_4/H_4
$,
where
\[
 F_4 = -{x}^{8}{y}^{4}+3\,{x}^{5}{y}^{3}+4\,{x}^{4}{y}^{2}-
x \left( 3+4\,x \right) y+1,
\]
and
\ifthenelse{\equal{\two}{true}} {
\begin{align*}
 H_4 = 
& {x}^{12}{y}^{6}-{x}^{8} \left( -x+2\,{x}^{2}+1 \right) {y}^{5} \\
& -{x}^{5} \left( 6\,{x}^{2}+2\,x+1+9\,{x}^{3}
 \right) {y}^{4} \\
& +2\,{x}^{3} \left( 13\,{x}^{2}+5\,x+1+4\,{x}^{3}
 \right) {y}^{3} \\
& +x \left( -6\,x-1+7\,{x}^{3}-6\,{x}^{2} \right) {y}^{2} \\
& - \left( x+1 \right)  \left( 6\,x+1 \right) y+1 .
\end{align*}
} {
\begin{multline*}
 H_4 = 
{x}^{12}{y}^{6}-{x}^{8} \left( -x+2\,{x}^{2}+1 \right) {y}^{5} %\\
-{x}^{5} \left( 6\,{x}^{2}+2\,x+1+9\,{x}^{3}
 \right) {y}^{4} \\
+2\,{x}^{3} \left( 13\,{x}^{2}+5\,x+1+4\,{x}^{3}
 \right) {y}^{3} %\\
+x \left( -6\,x-1+7\,{x}^{3}-6\,{x}^{2} \right) {y}^{2} %\\
- \left( x+1 \right)  \left( 6\,x+1 \right) y+1 .
\end{multline*}
}
For $n=5$,
$
 {G_5^c}=F_5/H_5
$,
where
\ifthenelse{\equal{\two}{true}} {
\begin{align*}
  F_5 = 
  & -{x}^{15}{y}^{6}+2\,{x}^{11} \left( -2+x \right) {y}
  ^{5} \\
  & +{x}^{8} \left( 8\,{x}^{2}+2+11\,x \right) {y}^{4}\\
  & +2\,{x}^{5}
 \left( 7\,{x}^{2}+3+8\,x \right) {y}^{3} \\
  & -{x}^{3} \left( 2-x+8\,{x}^{2} \right) {y}^{2} \\
  & -2\,x \left( 2+5\,x \right) y+1,
\end{align*}
} {
\begin{multline*}
  F_5 = -{x}^{15}{y}^{6}+2\,{x}^{11} \left( -2+x \right) {y}
  ^{5}+{x}^{8} \left( 8\,{x}^{2}+2+11\,x \right) {y}^{4}\\
  +2\,{x}^{5}
 \left( 7\,{x}^{2}+3+8\,x \right) {y}^{3}-{x}^{3} \left( 2-x+8\,{x}^{2
} \right) {y}^{2} %\\
 -2\,x \left( 2+5\,x \right) y+1,
\end{multline*}
}
and
\ifthenelse{\equal{\two}{true}} {
\begin{align*}
H_5 = 
&
{x}^{20}{y}^{8}+{x}^{15} \left( 3\,{x}^{2}-x+1 \right) {y}^{7} \\
&
-{x}^{11} \left( 19\,{x}^{4}
+11\,{x}^{3}+7\,{x}^{2}+2\,x+1 \right) {y}^{6} \\
&
-{x}^{8} \left( 2\,{x}^{4}+65\,{x}^{3}+39\,{x}^{2}+11\,x+2 \right) {y}^{5} \\
&
+{x}^{5} \left( 41\,
{x}^{5}+95\,{x}^{4}+39\,{x}^{3}-9\,{x}^{2}-6\,x-1 \right) {y}^{4} \\
&
+{x}^
{3} \left( 34\,{x}^{4}+85\,{x}^{3}+69\,{x}^{2}+19\,x+2 \right) {y}^{3} \\
&
-x \left( 19\,{x}^{4}+19\,{x}^{3}+27\,{x}^{2}+10\,x+1 \right) {y}^{2} \\
&
-
 \left( 15\,{x}^{2}+9\,x+1 \right) y+1 .
\end{align*}
} {
\begin{multline*}
H_5 = 
{x}^{20}{y}^{8}+{x}^
{15} \left( 3\,{x}^{2}-x+1 \right) {y}^{7}
-{x}^{11} \left( 19\,{x}^{4}
+11\,{x}^{3}+7\,{x}^{2}+2\,x+1 \right) {y}^{6} \\
-{x}^{8} \left( 2\,{x}^{
4}+65\,{x}^{3}+39\,{x}^{2}+11\,x+2 \right) {y}^{5} %\\
+{x}^{5} \left( 41\,
{x}^{5}+95\,{x}^{4}+39\,{x}^{3}-9\,{x}^{2}-6\,x-1 \right) {y}^{4} \\
+{x}^
{3} \left( 34\,{x}^{4}+85\,{x}^{3}+69\,{x}^{2}+19\,x+2 \right) {y}^{3} %\\
-x \left( 19\,{x}^{4}+19\,{x}^{3}+27\,{x}^{2}+10\,x+1 \right) {y}^{2} \\
-
 \left( 15\,{x}^{2}+9\,x+1 \right) y+1 .
\end{multline*}
}

\subsection{Lattices with free boundaries}
For $n=2$,
\[
G_2^{\text{fb}} = 
{\frac {1-xy}{{x}^{3}{y}^{3}-x{y}^{2} - \left( 2\,x+1 \right) y+1}} .
\]
For $n=3$,
$
 G_3^{\text{fb}} = F_3^{\text{fb}}/H_3^{\text{fb}}
$,
where
\[
 F_3^{\text{fb}} = 
 {x}^{6}{y}^{4}+{x}^{4}{y}^{3}-2\,{x}^{2} \left( 1+x \right) {y
}^{2}-xy+1,
\]
and
\ifthenelse{\equal{\two}{true}} {
\begin{align*}
 H_3^{\text{fb}} = 
 & -{x}^{9}{y}^{6}+{x}^{6} \left( x-1 \right) {y}^{5} \\
 & +{x}^{4}
 \left( 5\,{x}^{2}+3\,x+2 \right) {y}^{4} 
  +{x}^{2} \left( 2\,x+1
 \right)  \left( x-1 \right) {y}^{3} \\
 &  -x \left( 1+x \right)  \left( 5\,x
 +2 \right) {y}^{2}
  - \left( 1+3\,x \right) y+1 .
\end{align*}
} {
\begin{multline*}
 H_3^{\text{fb}} = -{x}^{9}{y}^{6}+{x}^{6} \left( x-1 \right) {y}^{5}+{x}^{4}
 \left( 5\,{x}^{2}+3\,x+2 \right) {y}^{4}+{x}^{2} \left( 2\,x+1
 \right)  \left( x-1 \right) {y}^{3}\\
 -x \left( 1+x \right)  \left( 5\,x
+2 \right) {y}^{2} - \left( 1+3\,x \right) y+1 .
\end{multline*}
}
For $n=4$, $G_4^{\text{fb}} = F_4^{\text{fb}}/H_4^{\text{fb}}$, with
\ifthenelse{\equal{\two}{true}} {
\begin{align*}
F_4^{\text{fb}} = 
& {x}^{14}{y}^{7}-2\,{x}^{11}{y}^{6}-{x}^{8} \left( 3\,x+5\,{x}^
{2}+3 \right) {y}^{5} \\
& +{x}^{6} \left( 2\,x+3 \right)  \left( 1+x
 \right) {y}^{4}\\
&  +{x}^{4} \left( 2\,x+3 \right)  \left( 1+3\,x \right)
{y}^{3} \\
& -3\,{x}^{2} \left( 1+3\,x+{x}^{2} \right) {y}^{2} \\
&-2\,x \left( 1
+x \right) y+1, 
\end{align*}
} {
\begin{multline*}
F_4^{\text{fb}} = 
{x}^{14}{y}^{7}-2\,{x}^{11}{y}^{6}-{x}^{8} \left( 3\,x+5\,{x}^
{2}+3 \right) {y}^{5}+{x}^{6} \left( 2\,x+3 \right)  \left( 1+x
 \right) {y}^{4}\\
 +{x}^{4} \left( 2\,x+3 \right)  \left( 1+3\,x \right)
{y}^{3}-3\,{x}^{2} \left( 1+3\,x+{x}^{2} \right) {y}^{2}-2\,x \left( 1
+x \right) y+1, 
\end{multline*}
}
and
\ifthenelse{\equal{\two}{true}} {
\begin{align*}
H_4^{\text{fb}} =
&
-{x}^{18}{y}^{9}+{x}^{14} \left( -x+{x}^{2}+1 \right)
{y}^{8} \\
&
+{x}^{11} \left( 2+9\,{x}^{3}+3\,x+9\,{x}^{2} \right) {y}^{7}\\
&
-{x}^{8} \left( 19\,{x}^{3}-1+6\,{x}^{2}+5\,{x}^{4} \right) {y}^{6} \\
&
-{x}^{6} \left( 29\,{x}^{2}+3+14\,x+24\,{x}^{3}+21\,{x}^{4} \right) {y}^{5}\\
&
+{x}^{4} \left( 41\,{x}^{2}+40\,{x}^{3}+9\,{x}^{4}+18\,x+3 \right) {y}^{4}\\
&
+{x}^{2} \left( 4\,{x}^{2}-1-4\,x+15\,{x}^{4}+27\,{x}^{3} \right)
{y}^{3}\\
&
-x \left( 5\,{x}^{3}+2+21\,{x}^{2}+13\,x \right) {y}^{2} \\
&
+ \left( -1-3\,{x}^{2}-5\,x \right) y+1 .
\end{align*}
} {
\begin{multline*}
H_4^{\text{fb}} =
-{x}^{18}{y}^{9}+{x}^{14} \left( -x+{x}^{2}+1 \right)
{y}^{8}+{x}^{11} \left( 2+9\,{x}^{3}+3\,x+9\,{x}^{2} \right) {y}^{7}\\
-{x}^{8} \left( 19\,{x}^{3}-1+6\,{x}^{2}+5\,{x}^{4} \right) {y}^{6}
-{x}^{6} \left( 29\,{x}^{2}+3+14\,x+24\,{x}^{3}+21\,{x}^{4} \right) {y}^{5}\\
+{x}^{4} \left( 41\,{x}^{2}+40\,{x}^{3}+9\,{x}^{4}+18\,x+3 \right) {y}
^{4}+{x}^{2} \left( 4\,{x}^{2}-1-4\,x+15\,{x}^{4}+27\,{x}^{3} \right)
{y}^{3}\\
-x \left( 5\,{x}^{3}+2+21\,{x}^{2}+13\,x \right) {y}^{2}
+
 \left( -1-3\,{x}^{2}-5\,x \right) y+1 .
\end{multline*}
}

\bibliography{md,md_1}

\begin{thebibliography}{10}
\expandafter\ifx\csname natexlab\endcsname\relax\def\natexlab#1{#1}\fi
\expandafter\ifx\csname bibnamefont\endcsname\relax
  \def\bibnamefont#1{#1}\fi
\expandafter\ifx\csname bibfnamefont\endcsname\relax
  \def\bibfnamefont#1{#1}\fi
\expandafter\ifx\csname citenamefont\endcsname\relax
  \def\citenamefont#1{#1}\fi
\expandafter\ifx\csname url\endcsname\relax
  \def\url#1{\texttt{#1}}\fi
\expandafter\ifx\csname urlprefix\endcsname\relax\def\urlprefix{URL }\fi
\providecommand{\bibinfo}[2]{#2}
\providecommand{\eprint}[2][]{\url{#2}}

\bibitem[{\citenamefont{Heilmann and Lieb}(1970)}]{Heilmann1970}
\bibinfo{author}{\bibfnamefont{O.~J.} \bibnamefont{Heilmann}} \bibnamefont{and}
  \bibinfo{author}{\bibfnamefont{E.~H.} \bibnamefont{Lieb}},
  \bibinfo{journal}{Phys. Rev. Lett.} \textbf{\bibinfo{volume}{24}},
  \bibinfo{pages}{1412} (\bibinfo{year}{1970}).

\bibitem[{\citenamefont{Kong}(2006{\natexlab{a}})}]{Kong2006c}
\bibinfo{author}{\bibfnamefont{Y.}~\bibnamefont{Kong}}, \bibinfo{journal}{Phys.
  Rev. E}  (\bibinfo{year}{2006}{\natexlab{a}}), \bibinfo{note}{to appear},
  \urlprefix\url{http://www.arxiv.org/cond-mat/0610690}.

\bibitem[{\citenamefont{Kasteleyn}(1961)}]{Kasteleyn1961}
\bibinfo{author}{\bibfnamefont{P.~W.} \bibnamefont{Kasteleyn}},
  \bibinfo{journal}{Physica} \textbf{\bibinfo{volume}{27}},
  \bibinfo{pages}{1209} (\bibinfo{year}{1961}).

\bibitem[{\citenamefont{Fisher}(1961)}]{Fisher1961}
\bibinfo{author}{\bibfnamefont{M.~E.} \bibnamefont{Fisher}},
  \bibinfo{journal}{Phys. Rev.} \textbf{\bibinfo{volume}{124}},
  \bibinfo{pages}{1664} (\bibinfo{year}{1961}).

\bibitem[{\citenamefont{Pemantle and Wilson}(2002)}]{Pemantle2002}
\bibinfo{author}{\bibfnamefont{R.}~\bibnamefont{Pemantle}} \bibnamefont{and}
  \bibinfo{author}{\bibfnamefont{M.~C.} \bibnamefont{Wilson}},
  \bibinfo{journal}{J. Comb. Theory Ser. A} \textbf{\bibinfo{volume}{97}},
  \bibinfo{pages}{129} (\bibinfo{year}{2002}).

\bibitem[{\citenamefont{Odlyzko}(1995)}]{Odlyzko95}
\bibinfo{author}{\bibfnamefont{A.~M.} \bibnamefont{Odlyzko}}, in
  \emph{\bibinfo{booktitle}{Handbook of Combinatorics}}, edited by
  \bibinfo{editor}{\bibfnamefont{R.~L.} \bibnamefont{Graham}},
  \bibinfo{editor}{\bibfnamefont{M.}~\bibnamefont{Gr{\"o}tschel}},
  \bibnamefont{and} \bibinfo{editor}{\bibfnamefont{L.}~\bibnamefont{Lovász}}
  (\bibinfo{publisher}{North-Holland}, \bibinfo{address}{Amsterdam},
  \bibinfo{year}{1995}), vol.~\bibinfo{volume}{2}, pp.
  \bibinfo{pages}{1063--1229}.

\bibitem[{\citenamefont{Kong}(1999)}]{Kong1999}
\bibinfo{author}{\bibfnamefont{Y.}~\bibnamefont{Kong}}, \bibinfo{journal}{J.
  Chem. Phys.} \textbf{\bibinfo{volume}{111}}, \bibinfo{pages}{4790}
  (\bibinfo{year}{1999}).

\bibitem[{\citenamefont{Kong}(2006{\natexlab{b}})}]{Kong2006}
\bibinfo{author}{\bibfnamefont{Y.}~\bibnamefont{Kong}}, \bibinfo{journal}{Phys.
  Rev. E} \textbf{\bibinfo{volume}{73}}, \bibinfo{pages}{016106}
  (\bibinfo{year}{2006}{\natexlab{b}}).

\bibitem[{\citenamefont{Kong}(2006{\natexlab{c}})}]{Kong2006b}
\bibinfo{author}{\bibfnamefont{Y.}~\bibnamefont{Kong}}, \bibinfo{journal}{Phys.
  Rev. E} \textbf{\bibinfo{volume}{74}}, \bibinfo{pages}{011102}
  (\bibinfo{year}{2006}{\natexlab{c}}).

\bibitem[{\citenamefont{Gaunt}(1969)}]{Gaunt1969}
\bibinfo{author}{\bibfnamefont{D.}~\bibnamefont{Gaunt}},
  \bibinfo{journal}{Phys. Rev.} \textbf{\bibinfo{volume}{179}},
  \bibinfo{pages}{174} (\bibinfo{year}{1969}).

\end{thebibliography}

\end{document}